\documentclass[submission,copyright,creativecommons]{eptcs}
\providecommand{\event}{GASCom 2024}

\usepackage{iftex}

\ifpdf
  \usepackage{underscore}         
  \usepackage[T1]{fontenc}        
\else
  \usepackage{breakurl}           
\fi

\usepackage{graphicx}
\usepackage{colortbl}
\usepackage{amsmath,amsfonts,amssymb}
\usepackage{amsthm}
\usepackage{pgf,tikz}
\usetikzlibrary{arrows,automata}
\usepackage{array}
\usepackage{tikz,tkz-tab}
\usepackage{blindtext}
\usepackage{hyperref}
\usepackage{stmaryrd}
\usepackage{booktabs}
\usepackage{url}
\usepackage{appendix}

\def\N{\mathbb{N}}
\def\Z{\mathbb{Z}}
\def\Q{\mathbb{Q}}

\def\A{\mathcal{A}}

\newcommand{\ind}{\mathbf 1}

\definecolor{lightgrey}{rgb}{0.9,0.9,0.9}
\definecolor{grey}{rgb}{0.7,0.7,0.7}
\definecolor{darkgrey}{rgb}{0.5,0.5,0.5}
\definecolor{lightblack}{rgb}{0.3,0.3,0.3}
\definecolor{lightblue}{rgb}{0.5,0.5,1}
\definecolor{lightred}{rgb}{1,0.5,0.5}
\definecolor{orange}{rgb}{1,0.8,0}
\definecolor{colorX}{rgb}{0,0.4,0}
\definecolor{colorY}{rgb}{0,0.7,0}
\definecolor{colorZ}{rgb}{0,1,0}

\definecolor{colorI}{rgb}{0,0,0}
\definecolor{colorA}{rgb}{0,0,1}
\definecolor{colorB}{rgb}{1,0,0}
\definecolor{colorC}{rgb}{0,0.4,0}
\definecolor{colorD}{rgb}{1,0.8,0}
\definecolor{colorE}{rgb}{0,1,0}
\definecolor{colorF}{rgb}{1,0.5,0.5}
\definecolor{colorG}{rgb}{0.3,0.65,1}
\definecolor{colorH}{rgb}{0.45,0.45,0.45}
\definecolor{colorJ}{rgb}{0.5,0,0.5}

\newcommand{\sq}[2] 
{\draw[lightgrey, fill=black] (#1,#2) -- (#1+1,#2) -- (#1+1, #2+1) -- (#1, #2+1) -- cycle;}

\newcommand{\sqB}[2] 
{\draw[lightgrey, fill=blue] (#1,#2) -- (#1+1,#2) -- (#1+1, #2+1) -- (#1, #2+1) -- cycle;}

\newcommand{\sqR}[2] 
{\draw[lightgrey, fill=red] (#1,#2) -- (#1+1,#2) -- (#1+1, #2+1) -- (#1, #2+1) -- cycle;}

\newcommand{\sqG}[2] 
{\draw[lightgrey, fill=darkgrey] (#1,#2) -- (#1+1,#2) -- (#1+1, #2+1) -- (#1, #2+1) -- cycle;}

\newcommand{\arrNE}[2] 
{\draw[thick, ->, lightblack] (#1+0.2,#2+0.2) -- (#1+0.8,#2+0.8) ;}

\newcommand{\arrNEblue}[2] 
{\draw[thick, ->, lightblue] (#1+0.2,#2+0.2) -- (#1+0.8,#2+0.8) ;}

\newcommand{\arrNWblue}[2] 
{\draw[thick, ->, lightblue] (#1+0.8,#2+0.2) -- (#1+0.2,#2+0.8) ;}

\newcommand{\arrNEBlue}[2] 
{\draw[thick, ->, blue] (#1+0.2,#2+0.2) -- (#1+0.8,#2+0.8) ;}

\newcommand{\arrNEX}[2] 
{\draw[thick, ->, colorX] (#1+0.2,#2+0.2) -- (#1+0.8,#2+0.8) ;}

\newcommand{\arrNWBlue}[2] 
{\draw[thick, ->, blue] (#1+0.8,#2+0.2) -- (#1+0.2,#2+0.8) ;}

\newcommand{\arrNEred}[2] 
{\draw[thick, ->, lightred] (#1+0.2,#2+0.2) -- (#1+0.8,#2+0.8) ;}

\newcommand{\arrNEgrey}[2] 
{\draw[thick, ->, darkgrey] (#1+0.2,#2+0.2) -- (#1+0.8,#2+0.8) ;}

\newcommand{\wall}[2] 
{\draw[thick, lightblack] (#1+0.5,#2+0.1) -- (#1+0.5,#2+0.9) ;}

\newcommand{\wallG}[2] 
{\draw[thick, grey] (#1+0.5,#2+0.1) -- (#1+0.5,#2+0.9) ;}

\newcommand{\wallR}[2] 
{\draw[thick, lightred] (#1+0.5,#2+0.1) -- (#1+0.5,#2+0.9) ;}

\newcommand{\wallX}[2] 
{\draw[thick, colorX] (#1+0.5,#2+0.1) -- (#1+0.5,#2+0.9) ;}

\newcommand{\wallRed}[2] 
{\draw[thick, red] (#1+0.5,#2+0.1) -- (#1+0.5,#2+0.9) ;}

\newcommand{\arrNWgrey}[2] 
{\draw[thick, grey, ->] (#1+0.8,#2+0.2) -- (#1+0.2,#2+0.8) ;}

\newcommand{\arrNEEblue}[2] 
{\draw[thick, lightblue, ->] (#1+0.2,#2+0.3) -- (#1+0.8,#2+0.7) ;}

\newcommand{\arrNEQ}[2] 
{\draw[thick, ->, lightblack] (#1+0.2,#2+0.2) -- (#1+0.8,#2+0.8) ;}

\newcommand{\arrNWQ}[2] 
{\draw[thick, ->, lightblack] (#1+0.8,#2+0.2) -- (#1+0.2,#2+0.8) ;}

\newcommand{\wallQ}[2] 
{\draw[thick, lightblack] (#1+0.5,#2+0.1) -- (#1+0.5,#2+0.9) ;}

\newcommand{\sqQ}[2] 
{\draw[lightgrey, fill=black] (#1,#2) -- (#1+1,#2) -- (#1+1, #2+1) -- (#1, #2+1) -- cycle;}

\newcommand{\sqBQ}[2] 
{\draw[lightgrey, fill=blue] (#1,#2) -- (#1+1,#2) -- (#1+1, #2+1) -- (#1, #2+1) -- cycle;}

\newcommand{\sqRQ}[2] 
{\draw[lightgrey, fill=red] (#1,#2) -- (#1+1,#2) -- (#1+1, #2+1) -- (#1, #2+1) -- cycle;}

\newcommand{\sqCR}[2] 
{\draw[red] (#1,#2) -- (#1+1,#2) -- (#1+1, #2+1) -- (#1, #2+1) -- cycle;}

\newcommand{\arrNEEQ}[2] 
{\draw[thick, lightblack, ->] (#1+0.2,#2+0.3) -- (#1+0.8,#2+0.7) ;}

\newcommand{\arrNEEBlue}[2] 
{\draw[thick, blue, ->] (#1+0.2,#2+0.3) -- (#1+0.8,#2+0.7) ;}

\newcommand{\arrNWWQ}[2] 
{\draw[thick, lightblack, ->] (#1+0.8,#2+0.3) -- (#1+0.2,#2+0.7) ;}

\newcommand{\arrNEEblueQ}[2] 
{\draw[thick, lightblue, ->] (#1+0.2,#2+0.4) -- (#1+0.8,#2+0.6) ;}

\newcommand{\arrNEEBlueQ}[2] 
{\draw[thick, blue, ->] (#1+0.2,#2+0.4) -- (#1+0.8,#2+0.6) ;}

\newcommand{\wallGQ}[2] 
{\draw[thick, grey] (#1+0.5,#2+0.1) -- (#1+0.5,#2+0.9) ;}

\newcommand{\wallBQ}[2] 
{\draw[thick, blue] (#1+0.5,#2+0.1) -- (#1+0.5,#2+0.9) ;}

\newcommand{\wallRQ}[2] 
{\draw[thick, red] (#1+0.5,#2+0.1) -- (#1+0.5,#2+0.9) ;}

\newcommand{\arrNEgreyQ}[2] 
{\draw[thick, grey, ->] (#1+0.2,#2+0.2) -- (#1+0.8,#2+0.8) ;}

\newcommand{\arrNQ}[2] 
{\draw[thick, lightblack, ->] (#1+0.2,#2+0.1) -- (#1+0.8,#2+0.9) ;}

\newcommand{\arrNBlue}[2] 
{\draw[thick, blue, ->] (#1+0.5,#2+0.1) -- (#1+0.5,#2+0.9) ;}

\newcommand{\arrNredQ}[2] 
{\draw[thick, red, ->] (#1+0.3,#2+0.2) -- (#1+0.7,#2+0.8) ;}

\newcommand{\arrNEBlueQ}[2] 
{\draw[thick, blue, ->] (#1+0.3,#2+0.2) -- (#1+0.7,#2+0.8) ;}

\newcommand{\arrNblueQ}[2] 
{\draw[thick, lightblue, ->] (#1+0.5,#2+0.1) -- (#1+0.5,#2+0.9) ;}

\newcommand{\arrNyellowQ}[2] 
{\draw[thick, yellow, ->] (#1+0.5,#2+0.1) -- (#1+0.5,#2+0.9) ;}

\newcommand{\arrNEredQ}[2] 
{\draw[thick, red, ->] (#1+0.2,#2+0.2) -- (#1+0.8,#2+0.8) ;}

\newcommand{\arrNEblueQ}[2] 
{\draw[thick, blue, ->] (#1+0.2,#2+0.2) -- (#1+0.8,#2+0.8) ;}

\newcommand{\arrNEblackQ}[2] 
{\draw[thick, black, ->] (#1+0.2,#2+0.2) -- (#1+0.8,#2+0.8) ;}

\newcommand{\wallBlue}[2] 
{\draw[thick, blue] (#1+0.5,#2+0.1) -- (#1+0.5,#2+0.9) ;}

\newcommand{\arrNWWcolorY}[2] 
{\draw[thick, ->, colorY] (#1+0.7,#2+0.3) -- (#1+0.3,#2+0.7) ;}

\newcommand{\arrNERed}[2] 
{\draw[thick, ->, red] (#1+0.2,#2+0.2) -- (#1+0.8,#2+0.8) ;}

\newcommand{\arrNWRed}[2] 
{\draw[thick, ->, red] (#1+0.8,#2+0.2) -- (#1+0.2,#2+0.8) ;}

\newcommand{\arrNW}[2] 
{\draw[thick, ->, lightblack] (#1+0.8,#2+0.2) -- (#1+0.2,#2+0.8) ;}

\newcounter{num}

\theoremstyle{definition}
\newtheorem{theorem}[num]{Theorem}
\newtheorem{Def}[num]{Definition}

\title{Construction of Minkowski Sums by Cellular Automata}
\author{Pierre-Adrien Tahay
\institute{Université de Lorraine, Loria, UMR 7503, F-54506 Vandœuvre-lès-Nancy, France}
\email{pierre-adrien.tahay@univ-lorraine.fr}
}
\def\titlerunning{Construction of Minkowski Sums by Cellular Automata}
\def\authorrunning{P.-A. Tahay}
\begin{document}
\maketitle

\begin{abstract}
We give a construction in a column of a one-dimensional cellular automaton of the Minkowski sum of two sets which can themselves occur in columns of cellular automata. It enables us to obtain another construction of the set of integers that are sums of three squares, answering a question by the same author in~\cite{tahay}. 
\end{abstract}

\section{Introduction}

A one-dimensional \emph{cellular automaton} (CA) is a dynamical system $(\A^{\Z}, F)$, where $\A$ is a finite set, and where the map $F:\A^{\Z}\rightarrow\A^{\Z}$ is defined by a local
rule acting uniformly and synchronously on the configuration space. More precisely, there exists an integer $r\geq 0$ called the \emph{radius} of the CA, and a \emph{local rule} $f:\A^{2r+1}\rightarrow\A$ such that \[F(x)_k=f((x_{k+i})_{-r\leq i\leq r}), \textrm{  for all  } x\in \A^{\Z}, \textrm{  and for all  }k\in\Z. \]
By the Curtis-Hedlund-Lyndon theorem, a map $F:\A^{\Z}\rightarrow\A^{\Z}$ is a CA if and only if it is continuous with respect to the product topology, and it commutes with the shift map $\sigma$ defined by 
\[\sigma(x)_k=x_{k-1}, \textrm{  for all  } x\in \A^{\Z},\textrm{  and for all  }k\in\Z.\]
A cellular automaton can be visualized by using a spacetime diagram consisting of a 2-dimensional grid where each cell contains an element of $\A$ and is represented by a space  and time coordinate.

The problem of representing a word (a sequence over a finite alphabet)  in a column of the spacetime diagram of a cellular automaton is an interesting one but relatively unexplored. One of the oldest results on the subject is the construction of the characteristic function of prime numbers by Fischer in 1965, using a cellular automaton with more than 30,000 states~\cite{fischer}. The number of states was considerably reduced by Korec in 1997 who provided another construction with only $11$ states~\cite{korec}. Following on from Fischer's work, Mazoyer and Terrier have established various results on words that can be realized as a column of a CA, which they call \emph{Fisher's constructible function}~\cite{mazoyer} (I think Fischer's name is misspelled as Fisher throughout their article, including when they cite~\cite{fischer} in the bibliography.)

Afterwards, several widely known words have been obtained as column of a CA. 
In 2015, Rowland and Yassawi, established an effective construction for all $p$-automatic sequences, for any prime number $p$, as column of a CA by using generating series and the theory of finite fields~\cite{rowland}. In 2022, Dolce and Tahay~\cite{dolce} obtained a construction for all Sturmian words having quadratic slope using the directive sequences and their ultimate periodicity.
Other constructions have been obtained by Marcovici, Stoll and Tahay in 2018 ~\cite{marcovici}, such as the characterisitic function of any polynomial $P\in\Q[X]$ of degree $d\geq 1$ with $P(\N)\subset \N$.

In this paper we establish the constructibility as a column of a CA of the Minkowski sum of two constructible sets. This generalizes the method used by the author for constructing the characteristic function of the set of integers that are sums of two squares from the construction of the characteristic function of the squares~\cite{tahay}.

\section{Signals in cellular automata}

In their paper~\cite{mazoyer}, Mazoyer and Terrier give some constructions such as the sum of two constructible functions, the linear combination of constructible functions, or constant-recursive sequences. They use \emph{signals} to obtain their various constructions. 
\begin{figure}[h!]
\begin{center}
\begin{tikzpicture}[scale=0.32]
\foreach \x in {0,...,17}{\draw[ultra thin, grey] (\x,0)--(\x,17) ;} 
\foreach \y in {0,...,17}{\draw[ultra thin, grey] (0,\y)--(17,\y);} 

\sq{6}{0}

\foreach \i in {1,...,16}{\wall{6}{\i}};
\foreach \i in {1,...,10} {\arrNEred{(6+\i}{(\i}} ;
\foreach \i in {1,...,5} {\arrNWWcolorY{(6-\i}{(3*\i}} ;
\foreach \i in {1,...,5} {\arrNEEblue{(2*\i+6}{(\i}} ;

\draw[-, ultra thick, black] (-4,-2)--(-3.5,-2) node[right, black]{vertical signal};
\draw[-, ultra thick, red] (5,-2)--(5.5,-2) node[right, black]{slope $1$};
\draw[ultra thick, colorY] (10,-2)--(10.5,-2) node[right, black]{slope $-3$};
\draw[ultra thick, blue] (16,-2)--(16.5,-2) node[right, black]{slope $1/2$};
\end{tikzpicture}

\caption{some instances of signals}
\end{center}
\end{figure}
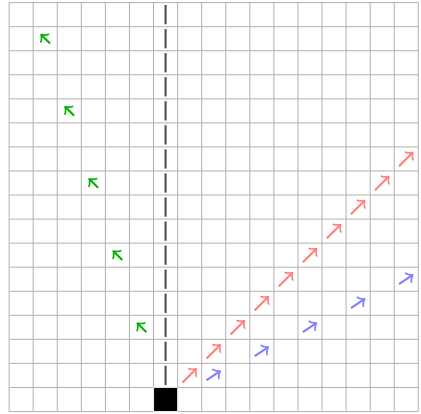

In the spacetime diagram of a CA, signals are a way to transmit information between two cells, by connecting two cells $(m,n)$ and $(m',n+t)$. The \emph{slope} of the signal is the number $\frac{t}{m'-m}$.

\section{Minkowski sums}

Let $A$ and $B$ be two sets. Recall that the \emph{Minkowski sum} of $A$ and $B$ is 
\[A+B = \{ a+b \ | \ a\in A, b\in B.\}\]

\begin{Def}
A set $A$ is called \emph{constructible by a CA} if the characteristic sequence $\ind_A$ of $A$ is obtained as a column of a CA.
\end{Def}

\begin{theorem}\label{MS}
Let $A$ and $B$ be two sets constructible by some cellular automata. Then the set $A+B$ is also constructible by a cellular automaton.
\end{theorem}

\begin{proof}
Let us call $G$ the cellular automaton that will build $\ind_{A+B}$. Let us call $F$ the cellular automaton that constructs $A$ in the left column of the CA. We compute the cellular automaton $\sigma \circ F$ and a signal of slope~$1$ in the diagonal of the spacetime diagram of $G$, so that we can mark the elements of $A$ on the diagonal. From these marked cells, we send vertical signals. Now we compute the set $B$ in the left column of $G$. From each element $b$ of $B$ in this left column we send a signal of slope~$1/2$. When this signal meets a vertical signal in a column of rank $a$ far any $a \in A$ we define a signal of slope~$-1/2$. When this signal meets the left column we are on the line of rank $b+a$. Since \[A+B=\bigcup\limits_{a \in A} \{a+b, \ b \in B\},\] the final set obtained in the left column is therefore $A+B$.

Note that signals with slope~$1/2$ can meet columns of rank $a$, with $a \in A$ below the vertical signal, in which case some elements $a+b$ could be missing. If there is only a finite number of these, then we define the first lines of the cellular automaton as initial conditions. If there is an infinite number of these, we change the signals of slopes~$1/2$ and $-1/2$ by signals of slopes $1$ and $-1$. Thus, we build en element of the form $b+2a$ with $b \in B$ and $a \in A$, but we can recover the elements of the form $a+b$ with $a \in A$ and $b \in B$ by using the cellular automaton $G^2$. 
\end{proof}

\section{Examples}

Let $S$ be the set of  squares, i.e. $S=\{n^2, n \in \N\}$.  So $S+S$ is the set of  integers that are sums of two squares and $S+(S+S)$ the set of the integers that are sums of three squares. We recall  below the constructions of $S$ and $S+S$ obtained in~\cite{tahay}. \\

\begin{figure}[h]
\quad\begin{minipage}[t]{.15\textwidth}
\begin{tikzpicture}[scale=0.25]

\foreach \x in {-1,...,10}{\draw[ultra thin, grey] (\x,0)--(\x,34) ;} 
\foreach \y in {0,...,34}{\draw[ultra thin, grey] (-1,\y)--(10,\y);} 

\foreach \x in {0,1,4,9,16,25}{\draw (-2,\x+0.5) node {{\small \x}} ;} 

\foreach \x in {0,...,33}{\wall{-1}{\x}} ; 

\foreach \x in {0,1,4,9,16,25}{\draw (0.5,\x+0.5) node {{\small 1}} ;} 
\foreach \x in {2,3,5,6,7,8,10,11,12,13,14,15,17,18,19,20,21,22,23,24,26,27,28,29,30,31,32,33}{\draw (0.5,\x+0.5) node {{\small 0}} ;}

\foreach \x in {0,1,4,9,16,25}{\sqCR{0}{\x} ;} 

\foreach \i in {0,...,2} {\wallX{2}{\i}} ;
\foreach \i in {3,...,6} {\wallX{3}{\i}} ;
\foreach \i in {7,...,12} {\wallX{4}{\i}} ;
\foreach \i in {13,...,20} {\wallX{5}{\i}} ;
\foreach \i in {21,...,30} {\wallX{6}{\i}} ;
\foreach \i in {31,...,33} {\wallX{7}{\i}} ;

\foreach \i in {1,...,1} {\arrNEBlue{\i}{1+\i}} ;
\foreach \i in {1,...,2} {\arrNEBlue{\i}{4+\i}} ;
\foreach \i in {1,...,3} {\arrNEBlue{\i}{9+\i}} ;
\foreach \i in {1,...,4} {\arrNEBlue{\i}{16+\i}} ;
\foreach \i in {1,...,5} {\arrNEBlue{\i}{25+\i}} ;

\foreach \i in {1,...,1} {\arrNWBlue{\i}{2+\i}} ;
\foreach \i in {1,...,2} {\arrNWBlue{3-\i}{6+\i}} ;
\foreach \i in {1,...,3} {\arrNWBlue{4-\i}{12+\i}} ;
\foreach \i in {1,...,4} {\arrNWBlue{5-\i}{20+\i}} ;
\foreach \i in {1,...,3} {\arrNWBlue{6-\i}{30+\i}} ;
\end{tikzpicture}
\end{minipage}
\hspace{1cm}
\begin{minipage}[t]{.80\textwidth}
\centering
\begin{tikzpicture}[scale=0.25]

\foreach \x in {-1,...,39}{\draw[ultra thin, grey] (\x,0)--(\x,33) ;} 
\foreach \y in {0,...,33}{\draw[ultra thin, grey] (-1,\y)--(39,\y);} 

\foreach \x in {0,1,2,4,5,8,9,10,13,16,17,18,20,25,26,29,32}{\draw (-2,\x+0.5) node {{\small \x}} ;} 

\foreach \x in {4,9,16,25,36}{\draw (\x+0.5,-1) node {{\small \x}} ;} 

\foreach \x in {0,...,32}{\wall{-1}{\x}} ; 

\foreach \x in {0,1,2,4,5,8,9,10,13,16,17,18,20,25,26,29,32}{\draw (0.5,\x+0.5) node {{\small 1}} ;} 
\foreach \x in {3,6,7,11,12,14,15,19,21,22,23,24,27,28,30,31}{\draw (0.5,\x+0.5) node {{\small 0}} ;}

\foreach \x in {0,1,4,9,16,25}{\sqCR{\x}{\x} ;} 

\foreach \i in {1,...,32} {\arrNEgreyQ{\i}{\i}} ;

\foreach \i in {0,...,3} {\wallX{2}{\i}} ;
\foreach \i in {4,...,7} {\wallX{3}{\i}} ;
\foreach \i in {8,...,13} {\wallX{4}{\i}} ;
\foreach \i in {14,...,21} {\wallX{5}{\i}} ;
\foreach \i in {22,...,31} {\wallX{6}{\i}} ;
\foreach \i in {32} {\wallX{7}{\i}} ;

\foreach \i in {1,...,1} {\arrNEBlue{\i}{1+\i}} ;
\foreach \i in {1,...,2} {\arrNEBlue{\i}{4+\i}} ;
\foreach \i in {1,...,3} {\arrNEBlue{\i}{9+\i}} ;
\foreach \i in {1,...,4} {\arrNEBlue{\i}{16+\i}} ;
\foreach \i in {1,...,5} {\arrNEBlue{\i}{25+\i}} ;

\foreach \i in {1,...,1} {\arrNWBlue{\i}{2+\i}} ;
\foreach \i in {1,...,2} {\arrNWBlue{3-\i}{6+\i}} ;
\foreach \i in {1,...,3} {\arrNWBlue{4-\i}{12+\i}} ;
\foreach \i in {1,...,4} {\arrNWBlue{5-\i}{20+\i}} ;
\foreach \i in {1,...,2} {\arrNWBlue{6-\i}{30+\i}} ;

\foreach \i in {1,...,1} {\arrNERed{\i}{2+\i}} ;
\foreach \i in {1,...,2} {\arrNERed{\i}{5+\i}} ;
\foreach \i in {1,...,3} {\arrNERed{\i}{10+\i}} ;
\foreach \i in {1,...,4} {\arrNERed{\i}{17+\i}} ;
\foreach \i in {1,...,5} {\arrNERed{\i}{26+\i}} ;

\foreach \i in {1,...,1} {\arrNWRed{\i}{3+\i}} ;
\foreach \i in {1,...,2} {\arrNWRed{3-\i}{7+\i}} ;
\foreach \i in {1,...,3} {\arrNWRed{4-\i}{13+\i}} ;
\foreach \i in {1,...,4} {\arrNWRed{5-\i}{21+\i}} ;
\foreach \i in {1,...,1} {\arrNWRed{6-\i}{31+\i}} ;

\foreach \i in {0,...,0} {\wallRQ{4}{4+\i}} ; 
\foreach \i in {0,...,0} {\wallRQ{9}{9+\i}} ;
\foreach \i in {0,...,6} {\wallRQ{16}{16+\i}} ;
\foreach \i in {0,...,6} {\wallRQ{25}{25+\i}} ;

\foreach \i in {0,...,0} {\arrNEEBlue{3+\i}{2+\i}} ;
\foreach \i in {0,...,1} {\arrNEEBlue{2*\i+6}{5+\i}} ;
\foreach \i in {0,...,2} {\arrNEEBlue{2*\i+11}{10+\i}} ;
\foreach \i in {0,...,3} {\arrNEEBlue{2*\i+18}{17+\i}} ;
\foreach \i in {0,...,4} {\arrNEEBlue{2*\i+27}{26+\i}} ;

\foreach \i in {0,...,29} {\wallBlue{4}{3+\i}} ;
\foreach \i in {0,...,25} {\wallBlue{9}{7+\i}} ;
\foreach \i in {0,...,19} {\wallBlue{16}{13+\i}} ;
\foreach \i in {0,...,11} {\wallBlue{25}{21+\i}} ;
\foreach \i in {0,...,1} {\wallBlue{36}{31+\i}} ;

\foreach \i in {-2,...,0} {\arrNEX{4+\i}{2+\i}} ;
\foreach \i in {0,...,3} {\arrNEX{6+\i}{3+\i}} ;
\foreach \i in {0,...,5} {\arrNEX{11+\i}{7+\i}} ;
\foreach \i in {0,...,7} {\arrNEX{18+\i}{13+\i}} ;
\foreach \i in {0,...,9} {\arrNEX{27+\i}{21+\i}} ;
\foreach \i in {0,...,0} {\arrNEX{38+\i}{31+\i}} ;

\foreach \i in {0,...,6} {\arrNEEQ{2*\i+2}{\i+5}} ;
\foreach \i in {0,...,13} {\arrNEEQ{2*\i+2}{\i+10}} ;
\foreach \i in {0,...,15} {\arrNEEQ{2*\i+2}{\i+17}} ;
\foreach \i in {0,...,6} {\arrNEEQ{2*\i+2}{\i+26}} ;

\foreach \i in {0,...,0} {\arrNWWQ{2*\i+2}{\i+7}} ;
\foreach \i in {0,...,0} {\arrNWWQ{2*\i+2}{\i+12}} ;
\foreach \i in {0,...,0} {\arrNWWQ{2*\i+2}{\i+19}} ;
\foreach \i in {0,...,0} {\arrNWWQ{2*\i+2}{\i+28}} ;

\foreach \i in {0,...,3} {\arrNWWQ{8-2*\i}{\i+9}} ;
\foreach \i in {0,...,3} {\arrNWWQ{8-2*\i}{\i+14}} ;
\foreach \i in {0,...,3} {\arrNWWQ{8-2*\i}{\i+21}} ;
\foreach \i in {0,...,2} {\arrNWWQ{8-2*\i}{\i+30}} ;

\foreach \i in {0,...,6} {\arrNWWQ{14-2*\i}{\i+18}} ;
\foreach \i in {0,...,6} {\arrNWWQ{14-2*\i}{\i+25}} ;

\foreach \i in {0,...,6} {\arrNWWQ{24-2*\i}{\i+22}} ;
\foreach \i in {0,...,3} {\arrNWWQ{24-2*\i}{\i+29}} ;

\end{tikzpicture}

\end{minipage}
\caption{CA for $S$ (left) and $S+S$ (right)}\label{fig:s12}
\end{figure}
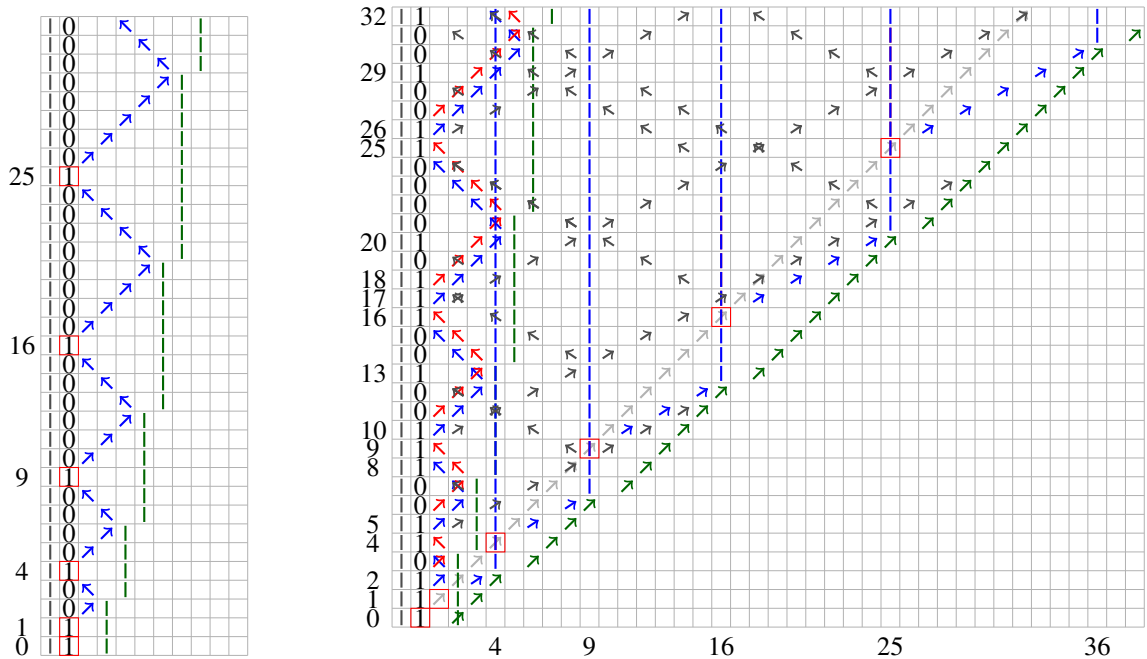

\bigskip
The construction of $S$ (see Figure~\ref{fig:s12}) is due to Delacourt, Poupet, Sablik and Theyssier~\cite{delacourt}. From a cell containing a $1$ in the left column, we send a signal of slope~$1$. When this signal meets a wall (vertical signal in green) we send a signal of slope~$-1$ and the wall is shifted one cell to the right and continues to spread vertically. The signal of slope~$-1$ marks the next square when it meets the left column.

For $S+S$ (see Figure~\ref{fig:s12}), we recall the construction previously obtained by the author in~\cite{tahay}. The principle is to start by constructing integers which are sums of two squares of the form $n^2+0^2$ and $n^2+1^2$ by using the same construction as for the squares. For the other elements of $S+S$ we mark the columns whose horizontal coordinate is a perfect square by using a method developped by Marcovici, Stoll and Tahay~\cite{marcovici} for the polynomial sequences. From each marked column we define a blue wall and from each perfect square in the left column we send a signal of slope $1/2$. When these signals meet a blue wall, we define a new signal a slope $-1/2$ which marks an element of $S+S$ when he meets the left column.

\begin{figure}[h]
\centering
\begin{tikzpicture}[scale=0.3]

\foreach \x in {-1,...,39}{\draw[ultra thin, grey] (\x,0)--(\x,33) ;} 
\foreach \y in {0,...,33}{\draw[ultra thin, grey] (-1,\y)--(39,\y);} 

\foreach \x in {0,1,2,3,4,5,6,8,9,10,11,12,13,14,16,17,18,19,20,21,22,24,25,26,27,29,30,32}{\draw (-2,\x+0.5) node {{\small \x}} ;} 

\foreach \x in {4,9,16,25,36}{\draw (\x+0.5,-1) node {{\small \x}} ;} 

\foreach \x in {0,...,32}{\wall{-1}{\x}} ; 

\foreach \x in {0,1,2,3,4,5,6,8,9,10,11,12,13,14,16,17,18,19,20,21,22,24,25,26,27,29,30,32}{\draw (0.5,\x+0.5) node {{\small 1}} ;} 
\foreach \x in {7,15,23,28,31}{\draw (0.5,\x+0.5) node {{\small 0}} ;}

\foreach \x in {0,1,4,9,16,25}{\sqCR{\x}{\x} ;}
\foreach \x in {0,1,2,4,5,8,9,10,13,16,17,18,20,25,26,29,32}{\sqCR{0}{\x} ;}  

\foreach \i in {1,...,32} {\arrNEgreyQ{\i}{\i}} ;






\foreach \i in {0,...,0} {\wallRQ{4}{4+\i}} ; 
\foreach \i in {0,...,0} {\wallRQ{9}{9+\i}} ;
\foreach \i in {0,...,6} {\wallRQ{16}{16+\i}} ;
\foreach \i in {0,...,6} {\wallRQ{25}{25+\i}} ;

\foreach \i in {0,...,0} {\arrNEEBlue{3+\i}{2+\i}} ;
\foreach \i in {0,...,1} {\arrNEEBlue{2*\i+6}{5+\i}} ;
\foreach \i in {0,...,2} {\arrNEEBlue{2*\i+11}{10+\i}} ;
\foreach \i in {0,...,3} {\arrNEEBlue{2*\i+18}{17+\i}} ;
\foreach \i in {0,...,4} {\arrNEEBlue{2*\i+27}{26+\i}} ;

\foreach \i in {0,...,29} {\wallBlue{4}{3+\i}} ;
\foreach \i in {0,...,25} {\wallBlue{9}{7+\i}} ;
\foreach \i in {0,...,19} {\wallBlue{16}{13+\i}} ;
\foreach \i in {0,...,11} {\wallBlue{25}{21+\i}} ;
\foreach \i in {0,...,1} {\wallBlue{36}{31+\i}} ;

\foreach \i in {-2,...,0} {\arrNEX{4+\i}{2+\i}} ;
\foreach \i in {0,...,3} {\arrNEX{6+\i}{3+\i}} ;
\foreach \i in {0,...,5} {\arrNEX{11+\i}{7+\i}} ;
\foreach \i in {0,...,7} {\arrNEX{18+\i}{13+\i}} ;
\foreach \i in {0,...,9} {\arrNEX{27+\i}{21+\i}} ;
\foreach \i in {0,...,0} {\arrNEX{38+\i}{31+\i}} ;

\foreach \i in {0,...,3} {\arrNEEQ{2*\i+2}{\i+3}} ;
\foreach \i in {0,...,6} {\arrNEEQ{2*\i+2}{\i+5}} ;
\foreach \i in {0,...,8} {\arrNEEQ{2*\i+2}{\i+6}} ;
\foreach \i in {0,...,11} {\arrNEEQ{2*\i+2}{\i+9}} ;
\foreach \i in {0,...,13} {\arrNEEQ{2*\i+2}{\i+10}} ;
\foreach \i in {0,...,14} {\arrNEEQ{2*\i+2}{\i+11}} ;
\foreach \i in {0,...,18} {\arrNEEQ{2*\i+2}{\i+14}} ;
\foreach \i in {0,...,15} {\arrNEEQ{2*\i+2}{\i+17}} ;
\foreach \i in {0,...,14} {\arrNEEQ{2*\i+2}{\i+18}} ;
\foreach \i in {0,...,13} {\arrNEEQ{2*\i+2}{\i+19}} ;
\foreach \i in {0,...,11} {\arrNEEQ{2*\i+2}{\i+21}} ;
\foreach \i in {0,...,6} {\arrNEEQ{2*\i+2}{\i+26}} ;
\foreach \i in {0,...,5} {\arrNEEQ{2*\i+2}{\i+27}} ;
\foreach \i in {0,...,2} {\arrNEEQ{2*\i+2}{\i+30}} ;

\foreach \i in {0,...,0} {\arrNWWQ{2*\i+2}{\i+5}} ;
\foreach \i in {0,...,0} {\arrNWWQ{2*\i+2}{\i+7}} ;
\foreach \i in {0,...,0} {\arrNWWQ{2*\i+2}{\i+8}} ;
\foreach \i in {0,...,0} {\arrNWWQ{2*\i+2}{\i+11}} ;
\foreach \i in {0,...,0} {\arrNWWQ{2*\i+2}{\i+12}} ;
\foreach \i in {0,...,0} {\arrNWWQ{2*\i+2}{\i+13}} ;
\foreach \i in {0,...,0} {\arrNWWQ{2*\i+2}{\i+16}} ;
\foreach \i in {0,...,0} {\arrNWWQ{2*\i+2}{\i+19}} ;
\foreach \i in {0,...,0} {\arrNWWQ{2*\i+2}{\i+20}} ;
\foreach \i in {0,...,0} {\arrNWWQ{2*\i+2}{\i+21}} ;
\foreach \i in {0,...,0} {\arrNWWQ{2*\i+2}{\i+23}} ;
\foreach \i in {0,...,0} {\arrNWWQ{2*\i+2}{\i+28}} ;
\foreach \i in {0,...,0} {\arrNWWQ{2*\i+2}{\i+29}} ;
\foreach \i in {0,...,0} {\arrNWWQ{2*\i+2}{\i+32}} ;

\foreach \i in {0,...,3} {\arrNWWQ{8-2*\i}{\i+7}} ;
\foreach \i in {0,...,3} {\arrNWWQ{8-2*\i}{\i+9}} ;
\foreach \i in {0,...,3} {\arrNWWQ{8-2*\i}{\i+10}} ;
\foreach \i in {0,...,3} {\arrNWWQ{8-2*\i}{\i+13}} ;
\foreach \i in {0,...,3} {\arrNWWQ{8-2*\i}{\i+14}} ;
\foreach \i in {0,...,3} {\arrNWWQ{8-2*\i}{\i+15}} ;
\foreach \i in {0,...,3} {\arrNWWQ{8-2*\i}{\i+18}} ;
\foreach \i in {0,...,3} {\arrNWWQ{8-2*\i}{\i+21}} ;
\foreach \i in {0,...,3} {\arrNWWQ{8-2*\i}{\i+22}} ;
\foreach \i in {0,...,3} {\arrNWWQ{8-2*\i}{\i+23}} ;
\foreach \i in {0,...,3} {\arrNWWQ{8-2*\i}{\i+25}} ;
\foreach \i in {0,...,2} {\arrNWWQ{8-2*\i}{\i+30}} ;
\foreach \i in {0,...,1} {\arrNWWQ{8-2*\i}{\i+31}} ;

\foreach \i in {0,...,6} {\arrNWWQ{14-2*\i}{\i+14}} ;
\foreach \i in {0,...,6} {\arrNWWQ{14-2*\i}{\i+17}} ;
\foreach \i in {0,...,6} {\arrNWWQ{14-2*\i}{\i+18}} ;
\foreach \i in {0,...,6} {\arrNWWQ{14-2*\i}{\i+19}} ;
\foreach \i in {0,...,6} {\arrNWWQ{14-2*\i}{\i+22}} ;
\foreach \i in {0,...,6} {\arrNWWQ{14-2*\i}{\i+25}} ;
\foreach \i in {0,...,6} {\arrNWWQ{14-2*\i}{\i+26}} ;
\foreach \i in {0,...,5} {\arrNWWQ{14-2*\i}{\i+27}} ;
\foreach \i in {0,...,3} {\arrNWWQ{14-2*\i}{\i+29}} ;

\foreach \i in {0,...,6} {\arrNWWQ{24-2*\i}{\i+21}} ;
\foreach \i in {0,...,6} {\arrNWWQ{24-2*\i}{\i+22}} ;
\foreach \i in {0,...,6} {\arrNWWQ{24-2*\i}{\i+23}} ;
\foreach \i in {0,...,6} {\arrNWWQ{24-2*\i}{\i+26}} ;
\foreach \i in {0,...,3} {\arrNWWQ{24-2*\i}{\i+29}} ;
\foreach \i in {0,...,2} {\arrNWWQ{24-2*\i}{\i+30}} ;
\foreach \i in {0,...,1} {\arrNWWQ{24-2*\i}{\i+31}} ;

\foreach \i in {0,...,0} {\arrNWWQ{34-2*\i}{\i+32}} ;

\end{tikzpicture}
\caption{Construction of $S+(S+S)$ by a cellular automaton}\label{fig:s3}
\end{figure}
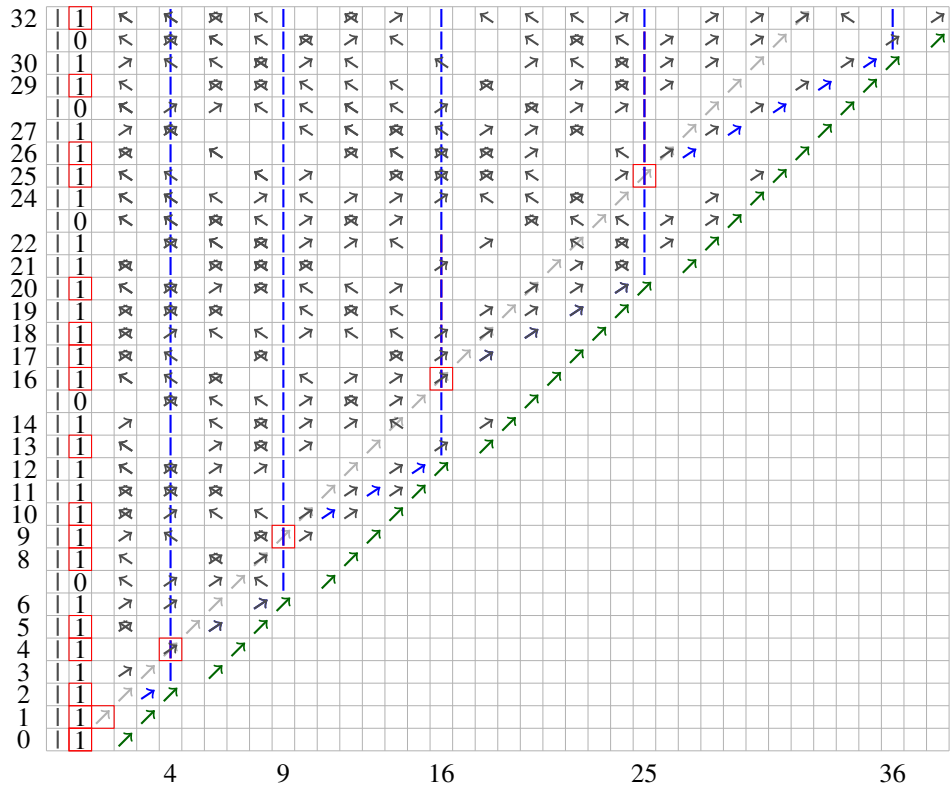

From these two constructions and Theorem~\ref{MS} we give a new construction of $S+(S+S)$ in Figure~\ref{fig:s3} using signals which answers the second open question in~\cite{tahay}. Note that in the figure, the first five lines are initial conditions.\\

\paragraph{Acknowledgment} The author thanks Guillaume Theyssier for asking him about the constructibility of Minkowski sums from those of the set of integers that are sums of two squares, that the author presented during a conference at CIRM on $29^{\text{th}}$ February 2024.

\nocite{*}
\bibliographystyle{eptcs}
\bibliography{Biblioeptcs}

\end{document}